\documentclass[conference]{IEEEtran}
\usepackage{cite}
\usepackage{graphicx}
\usepackage{mathtools}
\usepackage{subfig}

\newcommand{\comment}[1]{}

\begin{document}
%
\title{Programming Flows in Dense Mobile Environments: A Multi-user Diversity Perspective}

\author{\IEEEauthorblockN{Ioannis Gasparis}
\IEEEauthorblockA{UC Riverside\\
Riverside, CA\\
Email: ioannis.gasparis@email.ucr.edu}
\and
\IEEEauthorblockN{Ula\c{s} C. Kozat}
\IEEEauthorblockA{DOCOMO Innovations. Inc.\\
Palo Alto, CA\\
Email: kozat@docomoinnovations.com}
\and
\IEEEauthorblockN{M. O\~{g}uz Sunay}
\IEEEauthorblockA{Ozyegin University\\
Istanbul, Turkey\\
Email: oguz.sunay@ozyegin.edu.tr}}

\maketitle

\begin{abstract}
	The emergence of OpenFlow and Software Defined Networks brings new perspectives into how we design the next generation networks, where the number of base stations/access points as well as the devices per subscriber will be dramatically higher. In such dense environments, devices can communicate with each other directly and can attach to multiple base stations (or access points) for simultaneous data communication over multiple paths. This paper explores how networks can maximally enable this multi-path diversity through network programmability. In particular, we propose programmable flow clustering and set policies for inter-group as well as intra-group wireless scheduling. Further, we propose programmable demultiplexing of a single network flow onto multiple paths before the congestion areas and multiplexing them together post congestion areas. We show the benefits of such programmability first for legacy applications that cannot take advantage of multi-homing without such programmability. We then evaluate the benefits for smart applications that take advantage of multi-homing by  either opening multiple TCP connections over multiple paths or utilizing a transport protocol such as MP-TCP designed for supporting such environments.  More specifically, we built an emulation environment over Mininet for our experiments. Our evaluations using synthetic and trace driven channel models indicate that the proposed programmability in wireless scheduling and flow splitting can increase the throughput substantially for both the legacy applications and the current state of the art.

\end{abstract}


\section{Introduction}\label{sec:introduction}

Traditionally, densification and shortening the distance between the base stations and wireless subscribers have been the dominant factor in providing higher and higher rates for the end users.  We also see a trend with subscribers using multiple device types on the same shared subscription plans and many urban areas are covered by more than one base station and/or access points. E.g., an average adult American owns 4 digital devices 
and spends on average 34 hours per month using smart phones~\cite{devices-study}.
As the exponential demand increase continues, further densification is needed.  In the wireless industry, device to device discovery and communication are already being aggressively pursued (e.g., Bluetooth Low Energy and LTE Direct) and they will be increasingly more common in consumer products to enable seamless user experience. Already mobile platforms can provide support to simultaneously access multiple networks (e.g., LTE and WiFi) using new transport protocols such as MP-TCP. Although currently limited to few applications provided by the vendors or operators, such capability will be more commonly used as operators deploy heterogeneous networks (e.g., small cells coexist within macro cell coverage area and device to device relaying is enabled) and as applications demand higher bandwidth, lower delay or better reliability.

In parallel to these trends, Software Defined Networks (SDNs) and OpenFlow in particular generated a revived interest in building agile programmable networks. Taking advantage of multiple networks that cover the same area is for instance proposed as an application of programmable routing and access control \cite{kk13}.  Various proposals that target mobile SDN both in the radio access network and in the core network exist \cite{yazici14, MobileFlow, SoftCell, InnovationCellular, OpenRadio, SoftRan, SDNRF}. Although it is imperative to ask the question whether network programmability brings any benefits for legacy applications, it is also important to consider the same question for future applications that take advantage of the state of the art transport protocols. \comment{Our work attempts to answer these open issues in particularly interesting link aggregation scenarios for mobile operators.}

\begin{figure}[tb!]
     \begin{center}
        \subfloat[Small Cell Scenario]{%
            \label{fig:firstlayout}
           \includegraphics[width=0.3\columnwidth]{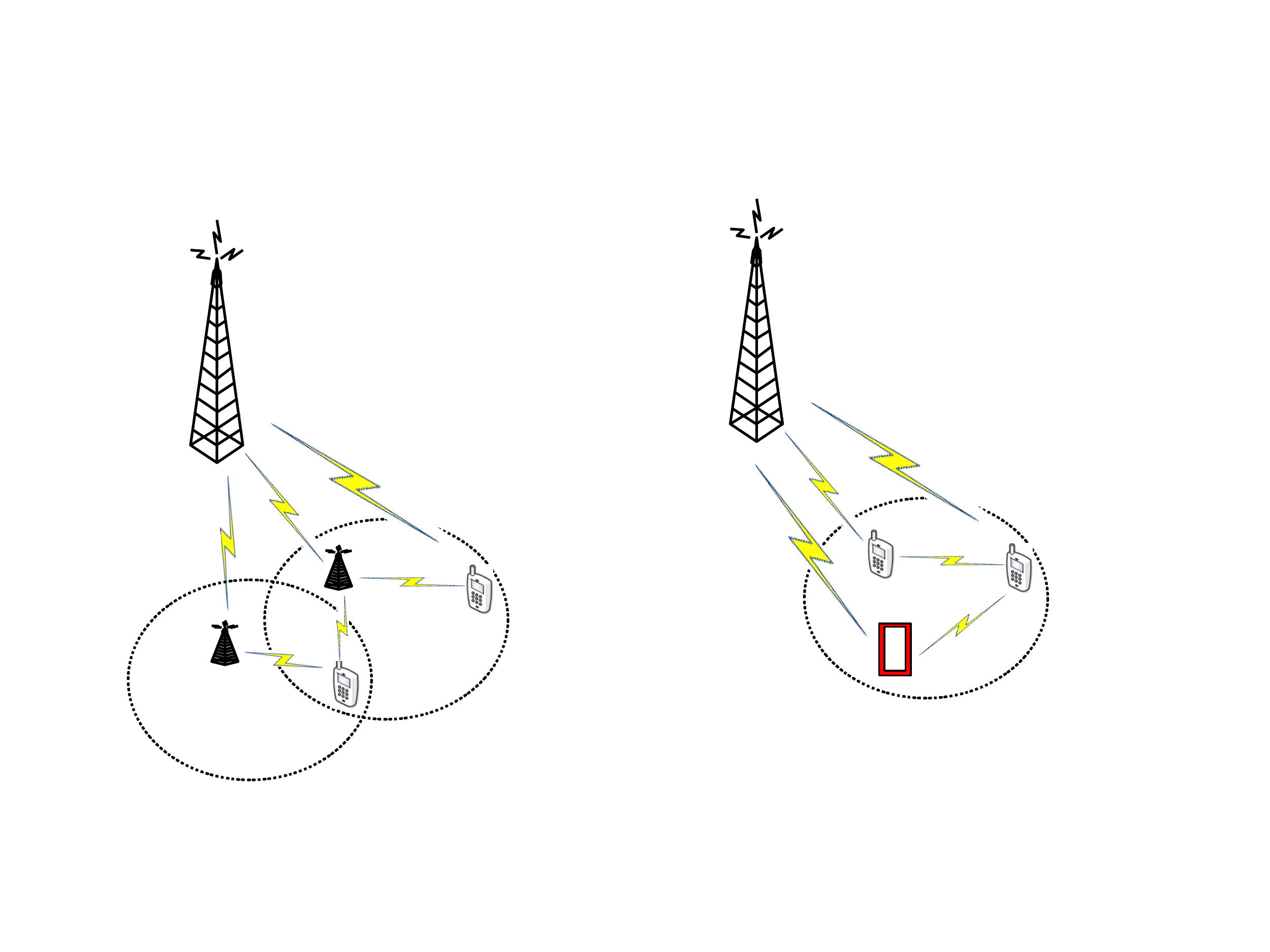}
        }%
        \subfloat[D2D Scenario]{%
           \label{fig:secondlayout}
           \includegraphics[width=0.3\columnwidth]{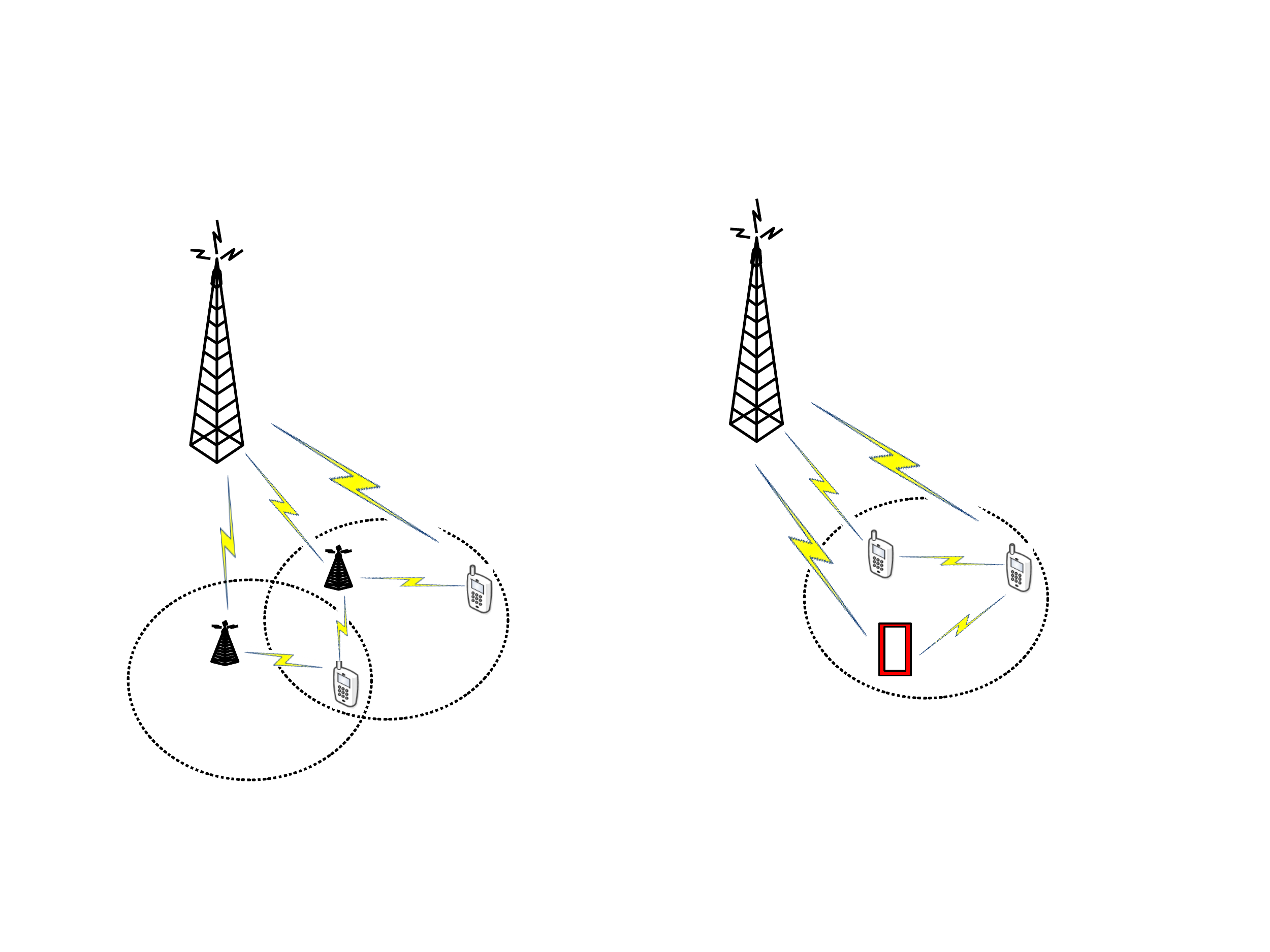}
        }    \end{center}
    \caption{%
        Shared wireless backhaul scenarios. 
     }%
   \label{fig:sharedbackhaul}
\end{figure}

Given the trends in network densification and SDN, we investigate the benefits of network programmability in particularly interesting link aggregation scenarios summarized in Fig.~\ref{fig:sharedbackhaul}. In the first use case (Fig.~\ref{fig:firstlayout}), the user is attached to a small cell and macro cell at the same time.  The backhaul for the small cell is wireless and it uses (i.e., contends for) the same spectrum resources as the end user (i.e., from base station point of view, small cell base station is another end user device that needs to be scheduled).  As a special case, the small cell base stations can be mobile routers mounted on top of trains, buses, cars, etc. In the second use case (Fig.~\ref{fig:secondlayout}), wireless subscriber carry multiple devices with device-to-device (D2D) links activated. While subscriber   consumes data on one device, she can utilize her other devices as relays. This use case can be extended easily to the case where devices that belong to different users are paired together and can perform wireless relaying for each other. The common point of these use cases is that there are either operator owned or personally owned wireless relays. We assume that the communication between the end host and wireless relay is orthogonal (i.e., non-interfering) with the communication between the end user/wireless relays and the macro cell base stations. 

The benefit of network programmability in the multi-path environment is that network operator knows how each flow is routed and which bottleneck link(s) they share. An obvious advantage of such a network intelligence happens when network knows which group of flows are for the same end host: Network controller can try to route each flow such that common bottleneck nodes/links are avoided. It becomes more interesting when the bottleneck nodes/links are unavoidable due to connectivity constraints. For instance, although multi-path routing is available in Fig.~\ref{fig:sharedbackhaul}, the bottleneck is the macro cell base station as relays and end users contend for the same resource blocks. Without any programmability in the wireless scheduler, a single scheduling policy (typically Proportional Fair Scheduling - PFS) runs and all the flows are treated with respect to the same notion of fairness. In the presence of a single end host, such scheduling would maximize the sum of logarithm of long term throughputs. However, as there is only one end host, the objective could instead be to maximize the sum rate (over the multi-path) for the end user. Therefore, the scheduler should have used opportunistic scheduling and serve the interface with the best instantaneous rate (referred to as Max C/I Scheduling).  In the presence of other end hosts that may or may not take advantage of multiple interfaces, we propose that there should be policy level programmability in the network where different network flows can be clustered together as a group. A network controller should be able to tell the base station what scheduling policy should be run for inter-group scheduling as well as for each intra-group scheduling. For instance, the inter-group scheduling can be set as PFS whereas for a particular group, intra-group scheduling can be set as Max C/I.  Even when multi-path aware transport protocols are used, as they  are not designed to take advantage of short-term channel opportunities, they should benefit from this level of network intelligence. The question is how much.

Besides programmability in the wireless scheduling, we also propose splitting a single flow at a network node into sub-flows (\emph{demux operation}) and combining them back together into a single flow (\emph{mux operation}) programmatically and transparent to the TCP layer and above. As we will present later in the paper, such dynamic flow demuxing and muxing not only is needed for legacy TCP applications that cannot take advantage of multi-homing, but it also performs slightly better than combining MP-TCP  and the proposed programmable scheduling.

The remainder of the paper is organized as follows: In Section~\ref{sec:related_work},  
we present some background and the related work. Our approach is detailed in Section~\ref{sec:approach}.
In Section~\ref{sec:implementation},  we describe our implementation over the Mininet framework while
in Section~\ref{sec:evaluation}, we validate our approach via experiments and discuss the results.
Our concluding remarks are presented in Section~\ref{sec:conclusion}.

\section{Related Work}\label{sec:related_work}

{\bf Mobile SDN \& Device-to-Device (D2D) Communications: } Mobile SDN is a relatively new area of research. An open wireless network infrastructure is recently proposed, where researchers 
can experiment with new services directly on production networks \cite{InnovationCellular}. Programmable wireless data plane \cite{OpenRadio}, application of core SDN ideas to mobile carrier networks \cite{MobileFlow},  redesign of the radio access and core networks again using SDN principles  \cite{SoftRan, SoftCell, yazici14} are all have been quite recent efforts in this domain. 

Although D2D on its own is investigated under the umbrella of ad hoc radio communications, delay tolerant networking, and hybrid wireless systems for quite a long time \cite{KozatThesis}, there is a renewed interest in D2D communications. This is mainly because D2D improves the spectrum and energy efficiency of the overall system through higher densification while increasing the throughput and end-to-end delay performance for the D2D links \cite{LinD2D}. 4G systems have already started studying device to device discovery and emergency communications in 3GPP Release 12 \cite{rel12}.  Various articles \cite{D2DCU, DND2D} describe the challenges of D2D communication in LTE-Advanced networks and propose solutions. In\cite{D2DVIDEO}, the authors present a scheme for increasing the
throughput of video files in a cellular network by leveraging the existence of D2D communication.  A new hierarchical control plane is proposed for 5G by designing an all-SDN network including the management of D2D communications \cite{yazici14}. Our paper is orthogonal to these works as we investigate the benefits of network programmability in link aggregation scenarios.  

{\bf Wireless Scheduling: } Wireless scheduling is maybe the most important network function in cellular systems as the wireless link is often the bottleneck. A comprehensive survey on the topic can be found in \cite{schedulingsurvey}. Almost all works on scheduling are designed with a particular objective and/or constraints in mind. For instance, maybe the most popular wireless scheduling in theory and practice consider proportional fairness as its target. In PFS, scheduler normalizes the instantaneous transmission rate of each user with respect to her own exponentially weighted average throughput. This normalized rate is used to pick the best user for the next resource block to be scheduled. It is fair in the sense that product of user throughputs is maximized. Another well-known scheduling policy is the opportunistic scheduling where the user with the highest current rate is always scheduled (Max C/I Scheduling). Although this maximizes the system throughput, it lacks fairness as users with relatively bad channel conditions would starve. Thus, in practice it is not used. As we will see later though Max C/I policy should be utilized in link aggregation scenarios. Yet another scheduler \cite{exprule} indicates the shortcomings of PFS and Max C/I in terms of queue stability. Programmable wireless scheduling is an emerging area. A recent work present a programmable packet scheduler that takes into the preferences of applications in scheduling over multiple network interfaces (e.g., preferences such as  "4G only", "WiFi only", or "both") \cite{kk13}. Different than the existing works, we propose that different scheduling policies should co-exist and picked by the network controller selectively as network state and capabilities evolve.   We also propose a specific programmability that can bundle flows together specifying inter-group and intra-group scheduling policies. Such group based scheduling notion exists in a limited scope. For instance, scheduling between two multicast groups is considered, where various fairness metrics across and inside the groups are investigated \cite{MSCHEDUL}.

{\bf Transport Layer Techniques: } There are many transport layer techniques that have 
been proposed for legacy and future applications in order to increase their 
quality. For single flows, techniques include Performance Enhancing Proxies~\cite{pep}, 
which are used in order to improve the TCP performance usually in a wireless environment or
SPDY~\cite{spdy}, which tries to reduce the load time of a web page by multiplexing and prioritizing the transfer of a web page into one flow.
P2P applications or video services such as Youtube and Netflix use multiple flows to fetch video segments in order to 
improve their reliability and performance. Multipath TCP~\cite{mptcp} tries to leverage the 
proliferation of the available network interfaces of a device by allowing a single data flow
to be split along different network paths. These solutions put more intelligence on the end devices and may in some cases replace the need for network intelligence (including programmability). Thus, one of the main objectives of this work is to understand how such end to end solutions perform with and without network programmability.

\section{Programmable Wireless Networks}\label{sec:approach}
\begin{figure}[!t]
	\centering
	\includegraphics[width=0.7\columnwidth]{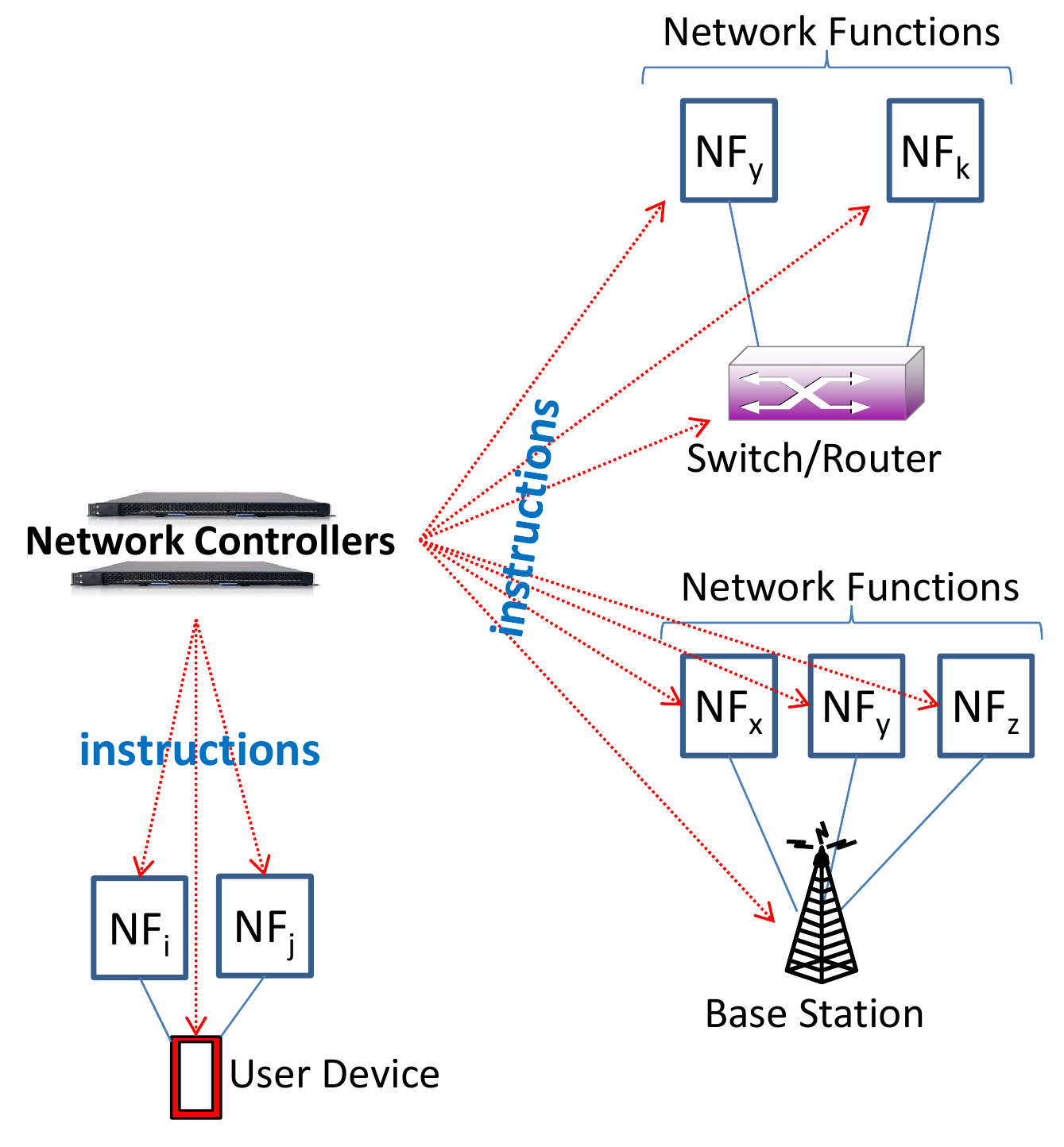}
	\vspace{-0.1in}
	\caption{Programmable Mobile Network Architecture}
	\label{fig:prognet} 
	\vspace{-0.13in}
\end{figure}

The model we adopt for wireless network programmability resonates with the proposal in \cite{yazici14}. As depicted in Fig.~\ref{fig:prognet}, programmability and network control are envisioned to be pervasive spanning user devices, base stations/access points, switches, and servers. Not only the forwarding tables, but also the network functions collocated with forwarding elements are considered to be programmable. These network functions (NFs) can be controller applications that read local state and take localized actions, but they can be also data plane functions that actually process the user payload requiring them to be physically close to the forwarding elements. Network controllers (which are distributed in a hierarchical way) push NFs onto the forwarding elements, program or configure the installed NFs, and install the forwarding rules into the flow tables of the forwarding elements. The controllers utilize forwarding table rules to selectively steer network traffic through a pipeline of network paths and NFs.  We present three programmable NFs to enable multi-path diversity in the core network and the wireless edge: Flow demultiplexer (Demux), flow multiplexer (Mux), and wireless scheduler.

\subsection{Programmable Flow Demultiplexer}
Splitting flows onto multiple paths for maximum utilization of network capacity is applied in many different contexts. With devices supporting multiple wireless links, SDN era provides new opportunities. As the wireless conditions might be greatly different for different users (even under the same set of base stations or access points), splitting user flows over multiple paths and wireless links cannot be a fixed policy that applies uniformly well to every user all the time. Furthermore, every time passing through the same network function when there are no significant gains adds to the system latency and reduces overall network throughput. Thus, identifying the right set of network conditions and selectively employing multi-path techniques only to the user flows that benefit from it is quite important. For instance, in Fig.~\ref{fig:secondlayout}, if the devices within the same proximity observe mostly good or highly correlated channels from the same base station, one should not expect significant gains. One can even observe performance loss due to the additional overheads in scheduling and forwarding. In contrast, if there is enough channel quality fluctuation and weak correlation, splitting a flow onto multi-path would be beneficial. \comment{ even for a two-second video segment requested from an HTTP server.}

The programmability for flow demultiplexer at least requires to identify (1) which flow to be split, (2) into how many subflows, (3) onto which links, and (4) the scheduling policy across subflows (e.g., round robin, WFQ, PFS, etc.). Naturally, the network controller that orchestrates the decision of multi-path diversity must place the flow demultiplexer onto the right forwarding element, program the demultiplexer, as well as program the routing tables such that the right flow is sent to the demux module and each subflow is forwarded to the right network interface.

\subsection{Programmable Flow Multiplexer}
Flows coming from multiple interfaces can be combined at the end device by the transport layer protocol such as TCP or by the application layer if they can handle the path delay variations and out of order packet delivery gracefully. Since many TCP variants adopt fast failure recovery feature, out of order packet delivery is a serious problem necessitating a flow multiplexer module at or before the end host. The placement of this function, in the link aggregation cases considered here, would be typically at the end host. However, if congestion is happening inside the core network or backhaul links, then Mux function can be placed on the switches or base stations.  

Programming a flow multiplexer amounts to (1) populating forwarding table entries for the subflows to steer them to the Mux function, (2) instructing Mux function about which subflows are to be multiplexed, (3) instructing the Mux function about multiplexing policy (e.g., in-order delivery, out-of-order delivery, multiplexing ratios, etc.), and (4) instructing the Mux function about buffering requirement for in-order packet delivery (e.g., hold packets for up to $T$ milliseconds if packets with smaller sequence numbers are missing). 

\subsection{Programmable Wireless Scheduler}
We divide the scheduling into two tiers for multi-path programmability. The network controller (NC) first groups flows/subflows $f_1, \hdots, f_N$ to be scheduled by a given base station into distinct clusters $C_1, \hdots, C_M$ and programs the scheduler. Then, NC specifies the scheduling policy for inter-cluster scheduling and the scheduling policy for intra-cluster scheduling. For instance, when inter-cluster scheduling policy is PFS, then scheduler assigns the next resource block to the cluster $C^*$ with the highest normalized rate, i.e., $C^{*} = \arg\max_{C_i} {\frac{ R(C_i) }{\overline{T}_{C_i} } }$. Here, $R(C_i)$ is the transmission rate to cluster $i$ and $\overline{T}_{C_i}$ is the average throughput of that cluster. If inter-cluster scheduling policy is maximum C/I, then 
$C^* = \arg\max_{C_i} R(C_i)$. 
Intra-scheduling policy determines which flow (or equivalently wireless link) is scheduled over the resource block assigned to $C_i$ and its transmission rate. If the policy is PFS, then $f_{i}^* = \arg\max_{f_j \in C_i} {\frac{R(f_i)}{\overline{T}_{f_i}} }$. Here, $R(f_i)$ is the current transmission rate  and average throughput $\overline{T}_{f_i}$ for $f_i$, respectively. Similarly, if the policy is Max C/I, then $f_{i}^* = \arg\max_{f_j \in C_i} {R(f_i)}$. The current transmission rate of cluster then becomes $R(C_i) = R(f_i^*)$. 

Multi-path diversity alters the classical notions of fairness across flows. For instance in the link aggregation scenarios in Fig.~\ref{fig:sharedbackhaul}, we do not need to be fair within the same cluster of flows unless fairness is important due to other layer considerations\comment{ (e.g., upper layer congestion protocol that requires periodic in-band signaling for performing multi-path optimization)}. Thus, although for inter-cluster scheduling we enforce fairness (e.g., proportional fairness), for intra-cluster scheduling we can pick to be opportunistic (e.g., Max C/I).

The combination of demux, mux and scheduling are sufficient to realize multi-path diversity gains. Clearly though, these add higher processing complexity, requiring them to be applied when the right set of link aggregation scenarios emerge (e.g., channel conditions are favorable to attain high performance gains). Next, we investigate this issue deeper.

\section{Implementation}\label{sec:implementation}
\subsection{Overview}\label{sec:implementation-overview}
We have proto-typed our programmable wireless network over  Mininet~\cite{mininet}, an open-source network emulator which
creates a network of virtual Linux hosts, SDN-enabled switches, 
OpenFlow controllers and links.  Mininet does not support wireless emulation, natively. Thus, we developed LTE-Emulator and D2D (e.g., WiFi) emulator to test various network topologies. We developed the wireless scheduler, Mux, and Demux as both standalone and integrated network functions (NFs). Each (stand alone or integrated) NF is attached as a Linux host to the switching fabric. Flows are steered towards each NF by programming the routing tables of switches. NFs capture packets above the link layer transparent to the TCP/IP stack and applications. Due to the overheads of emulation environment and our desire to come as close to the LTE rates as possible, we run our experiments over the integrated mode where the Demux and scheduling modules are integrated within the LTE-Emulator.  Our emulator is agnostic to the type of the network flow, meaning that the end hosts can 
run off-the-self applications (e.g., iperf) using an unmodified Linux network stack.  
In order to run a scenario, the user should specify an xml file having the 
appropriate parameters for the demux and scheduling policy as well as the parameters 
for the LTE and D2D channel. As it is the most important component, we provide more details on the LTE-Emulator.

\subsection{Design of LTE-Emulator}\label{sec:implementation-design}
\begin{figure}[!t]
	\centering
	\includegraphics[scale=0.30]{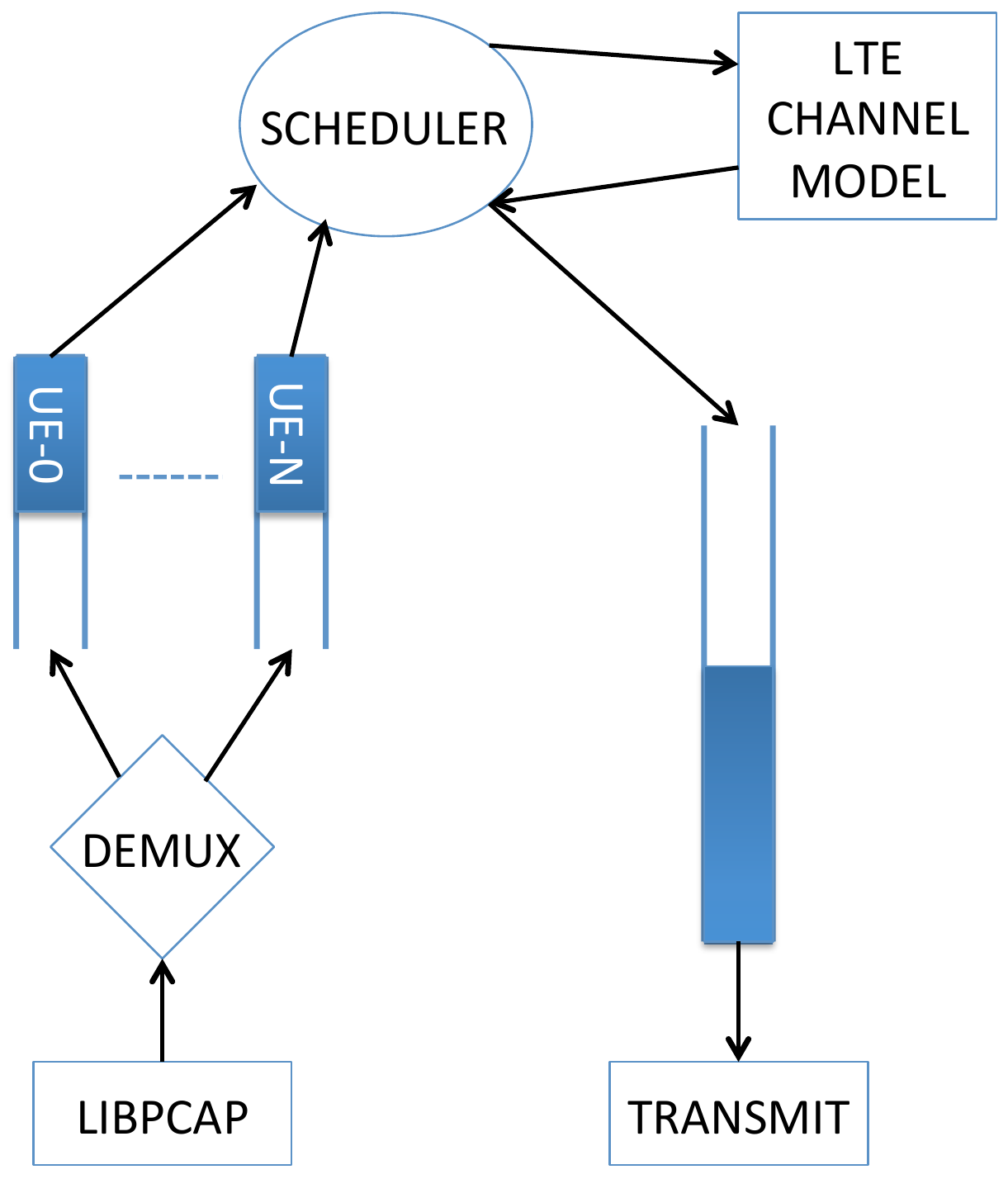}
	\vspace{-0in}
	\caption{LTE Emulator Design}
	\label{fig:emulator_design} 
	\vspace{-0.13in}
\end{figure}

The LTE-Emulator together with an Open vSwitch (OVS) emulates the programmable eNB (i.e., LTE base station). All the flows targeting a wireless host is forwarded by the OVS to the LTE-Emulator that has four components:  Network, 
demux, scheduler, and the LTE channel model (see Fig.~\ref{fig:emulator_design}). 

As LTE-Emulator is another host from the point of Mininet, the packets must be captured below the TCP/IP stack. The network component is responsible for capturing and passing the ethernet frames received from the switch to the Demux module.  To this end, we utilize the libpcap-1.6.1 library.

The demux module splits the ethernet frames into distinct buffers for individual LTE links based on a programmable buffer management policy. If a flow is to be split over multiple next hop nodes, its frames are copied to the corresponding buffers of the candidate next hops. Whenever scheduler pulls an LTE block from a particular buffer, demux module removes the same block from all the buffers. 

Scheduler implements the wireless scheduling policy configured by the network controller. It uses the wireless link states set by the LTE channel model to schedule wireless users. It also  determines how many bits should be transmitted for the scheduled user during the next scheduling interval (i.e., 1 ms in LTE) based on the state of the wireless link.  

The LTE channel model component runs the LTE channel models per UE and provides channel feedback to the scheduler. We emulate only the LTE downlink. In its current version, we have not implemented  radio link layer retransmission mechanisms nor the control signaling between eNBs and wireless devices. Hence, we run the LTE-Emulator only on eNB, and not on the wireless hosts. The uplink communication is switched through OVS without any involvement of NFs.

\section{Performance Evaluations}\label{sec:evaluation}
This section focuses on the impact of the scheduling policy (PFS vs. Max C/I)
on the user's throughput under different transport layer options and channel models. 

\subsection{Transport  Scenarios}\label{sec:eval-scenarios}
We consider the following four scenarios:

{\textbf{Single Flow:}} In this scenario, there is no link aggregation and TCP is used as the transport layer protocol. The end host is directly served by eNB. 

\begin{figure}[!t]
	\centering
	\subfloat[Demuxed TCP\label{fig:single-tcp}]{\includegraphics[width=0.25\columnwidth]{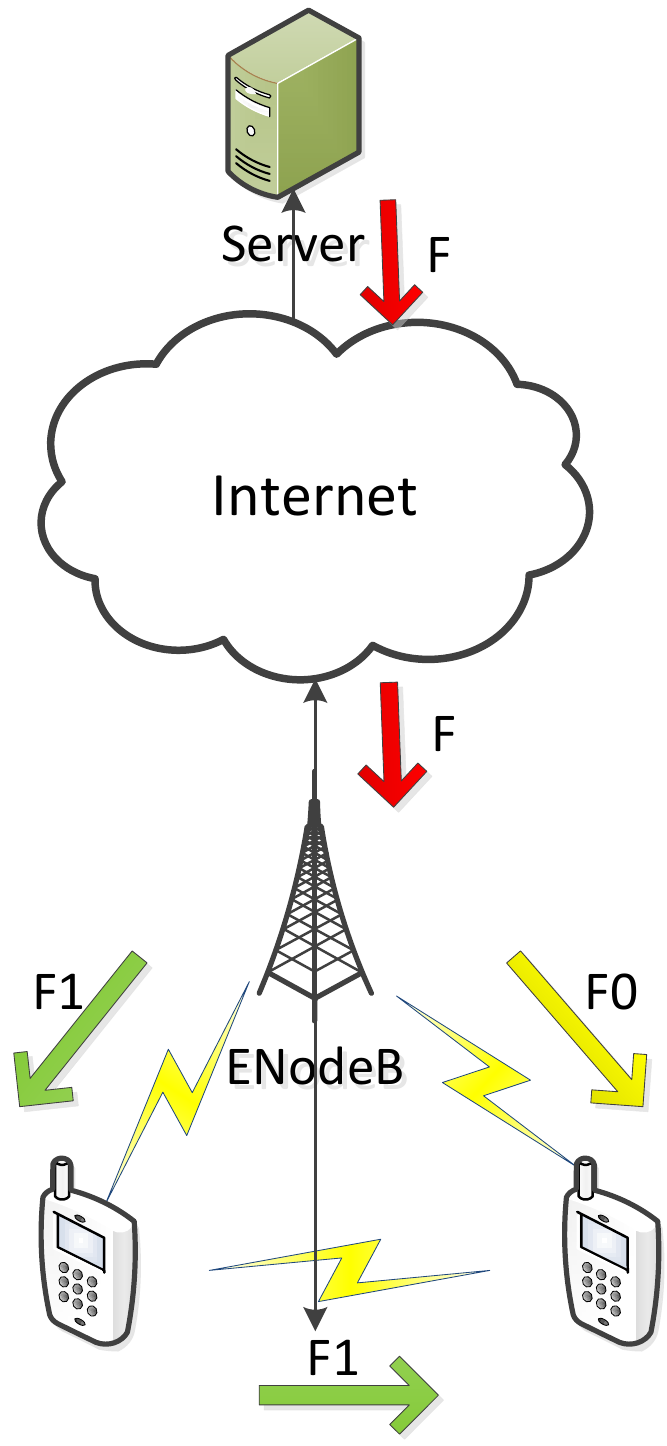}} \hspace{0.2cm}
	\subfloat[Multiple TCP\label{fig:multi-tcp}]{\includegraphics[width=0.25\columnwidth]{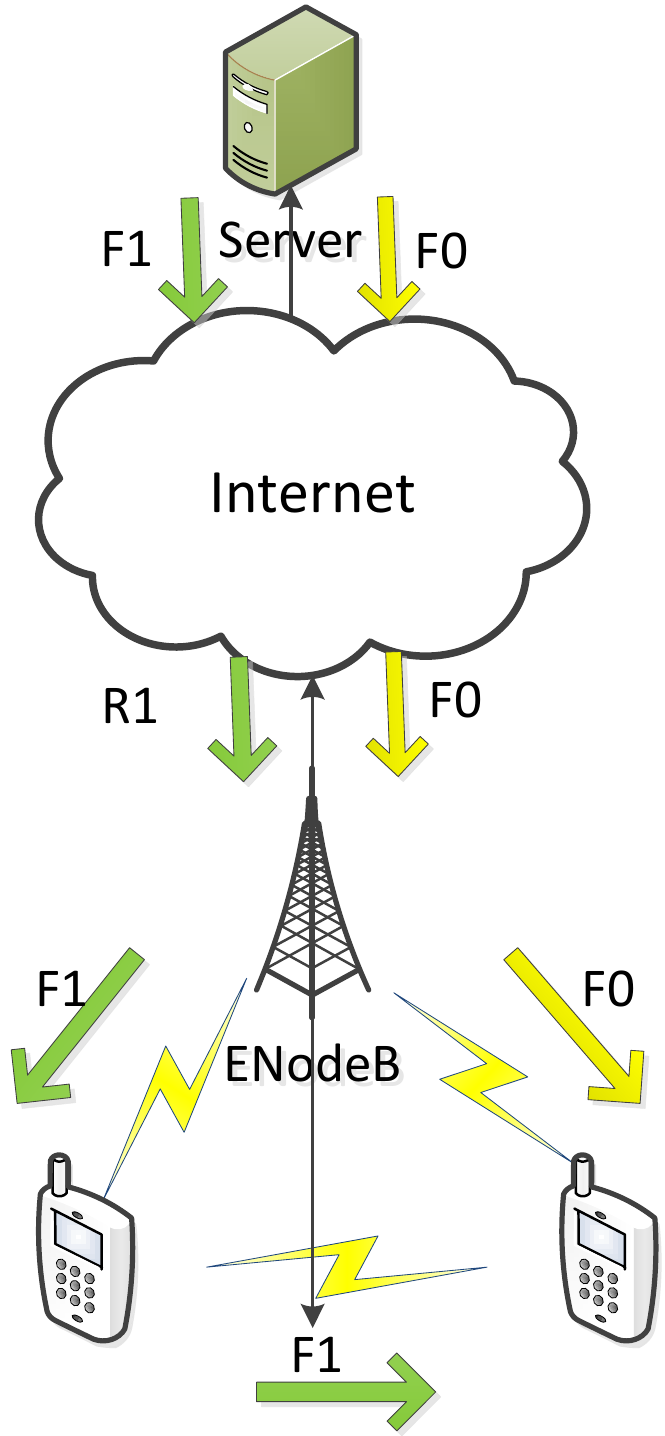}} \hspace{0.2cm}
	\subfloat[MP-TCP\label{fig:mptcp}]{\includegraphics[width=0.25\columnwidth]{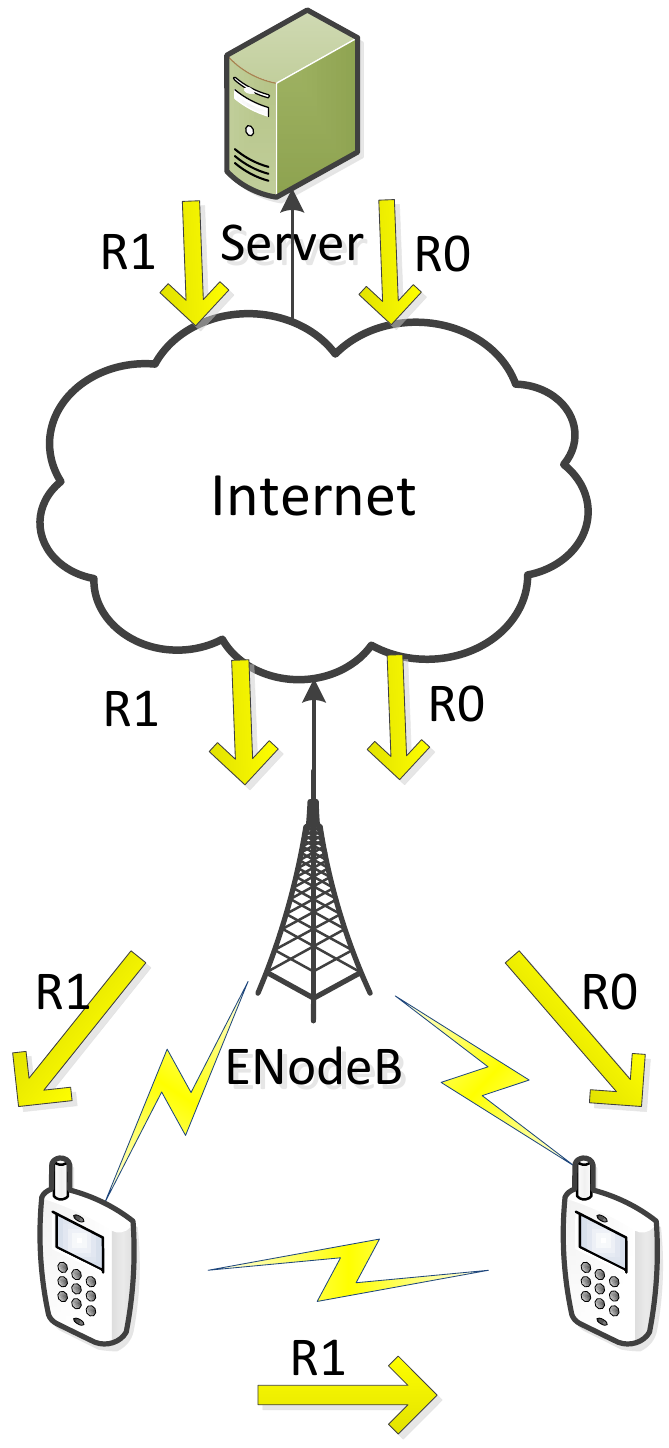}}
	\caption{Transport Scenarios}
	\label{fig:scenarios} 
	\vspace{-0.13in}
\end{figure}

{\textbf{Demuxed TCP Flow:}} This scenario is depicted in Fig.~\ref{fig:single-tcp}. Both the server and the end host use single TCP session for the communication (thus, there is no support for multi-homing at the transport layer). The flow is split at the base station via the Demux module into two sub-flows transparent to the end host applications. One flow is directly transmitted to the sink node and the other is  relayed through another LTE device. To enable relaying, NC programs the routing rules  in the relay node. 

{\textbf{Multiple TCP flows:}}  In this scenario (see Fig.~\ref{fig:multi-tcp}), the end host opens two separate TCP connections to the server (one connection over LTE link and the other connection over D2D link). Thus, the end host can stream two decoupled TCP flows into the network. In the absence of network programmability, the relay node acts in tethered mode, and from the base station point of view these flows are destined for two different LTE devices. The default fairness policy is applied to both flows. 

{\textbf{Multipath TCP (MP-TCP):}} In this scenario, the application utilizes MP-TCP. Upon detecting the availability of both LTE and D2D links (see Fig.~\ref{fig:mptcp}), MP-TCP starts using both interfaces and performs the multiplexing operation.  The main difference from the multiple TCP flow case is that the congestion windows of different flows are coupled together according to the congestion window resizing procedure of MP-TCP. MP-TCP tries to allocate bandwidth on each interface based on the relative round trip time values over each path. Such coupling results in better fairness against other users over the shared bottleneck links than the multiple TCP flow case.

 If network programming is available, then network controller can program the routing tables for multi-path routing and wireless relaying. In the absence of programmable scheduler, the default scheduling is set as PFS. With programmable scheduler, the two flows destined for the same end host are bundled together as a cluster and an intra-cluster scheduling policy is specified. We set the objective as to maximize the sum rate at the end host, therefore we utilize Max C/I as the intra-cluster scheduling policy. For inter-cluster scheduling, we apply proportional fairness.

Note that except for the single flow case,  all other transport layer solution can potentially benefit from network programmability. We will see in the results how MP-TCP and two TCP flow cases attain significantly better performance with programmability in the wireless scheduler.

\subsection{Channel Models}\label{sec:eval-channel}
For modeling the channel states, we use a simple Markovian model as well as real wireless traces. 
The Markovian model collapses the highest half of the LTE transmission rates into Good state and lower half of the LTE rates into 
Bad state. The transition probabilities from Bad state to Good state and from Good state to Bad state are denoted as $p$ and $q$, respectively. Once in a Good or Bad state, during each scheduling interval, the transmission rate of a wireless device is set uniformly random among the LTE rates represented by that state. Each wireless device has independent channel realizations from each other with different $p$ and $q$ parameters.

For real time traces, we use LTE channel measurement data collected in Tokyo, Japan under urban vehicular conditions. \comment{The trace files provide received signal strength from the serving cell as well as the neighboring cells, which allows us to compute signal to interference ratio and the corresponding channel quality indicator (CQI) values. CQI values are directly mapped onto LTE rates.} The traces provide signal strength and interference measurements directly acquired from the LTE chip on the smart phones. The measurements were logged every 1 second and averaged over few hundred milliseconds. In reality, wireless scheduler has feedback in the order of ten milliseconds, and hence track channel opportunities at a much faster time scale. To mitigate this shortcoming to some extent, we eliminated the traces that do not show significant channel fluctuations, as there is no multi-user diversity gain to be expected for such cases. In  our trace driven simulations, we assumed that the channel conditions for each user remains the same between measurement timestamps. Since D2D link must be in close proximity, to capture the location coherence for the end host and the relay node, we use the same trace file (i.e., collected from the same LTE device) but time shift the values for 2 measurement epochs.  For devices that do not have any D2D link, we use a distinct trace file.

\begin{figure}[!t]
	\centering
	\includegraphics[width=0.7\columnwidth]{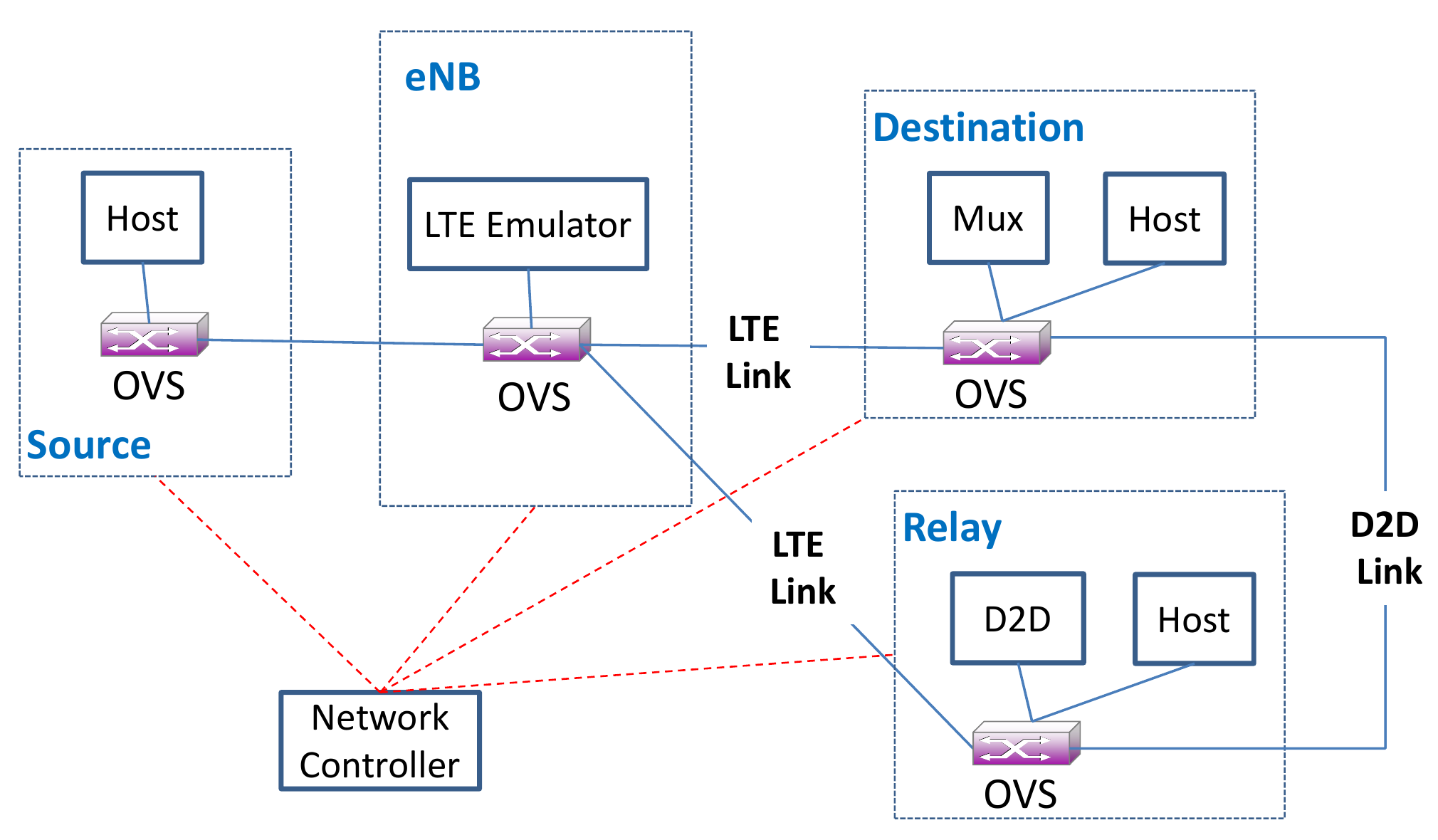}
	\caption{Mininet Topology}
	\label{fig:mininet-topology} 
	\vspace{-0.13in}
\end{figure}

\subsection{Experiments}\label{sec:eval-experiments}
We conducted extensive experiments using a PC Intel Core 2 Duo 2.66GHz processor with 2 GB RAM, Debian Jessie 32-bit kernel (3.11.10 kernel with mptcp) and running our LTE Emulator over the Mininet framework. 
For the initial set up, we used the Mininet topology in Fig.~\ref{fig:mininet-topology}. 
The source, destination and network functions run inside Mininet hosts and are
connected to the Open vSwitch (OVS) using 1 Gbps ethernet links.  Backhaul OVS switches are also interconnected with 1Gbps links. We assume that the bottleneck is the LTE, hence we emulate the D2D link using a stable 300 Mbps link, representative of Wi-Fi speeds.  LTE's theoretical maximum throughput is 75 Mbps, while the actual one that we can achieve is 65 Mbps due to the limitations in our emulation environment. 
 
\begin{figure}[!t]
	\centering
	\includegraphics[width=0.8\columnwidth]{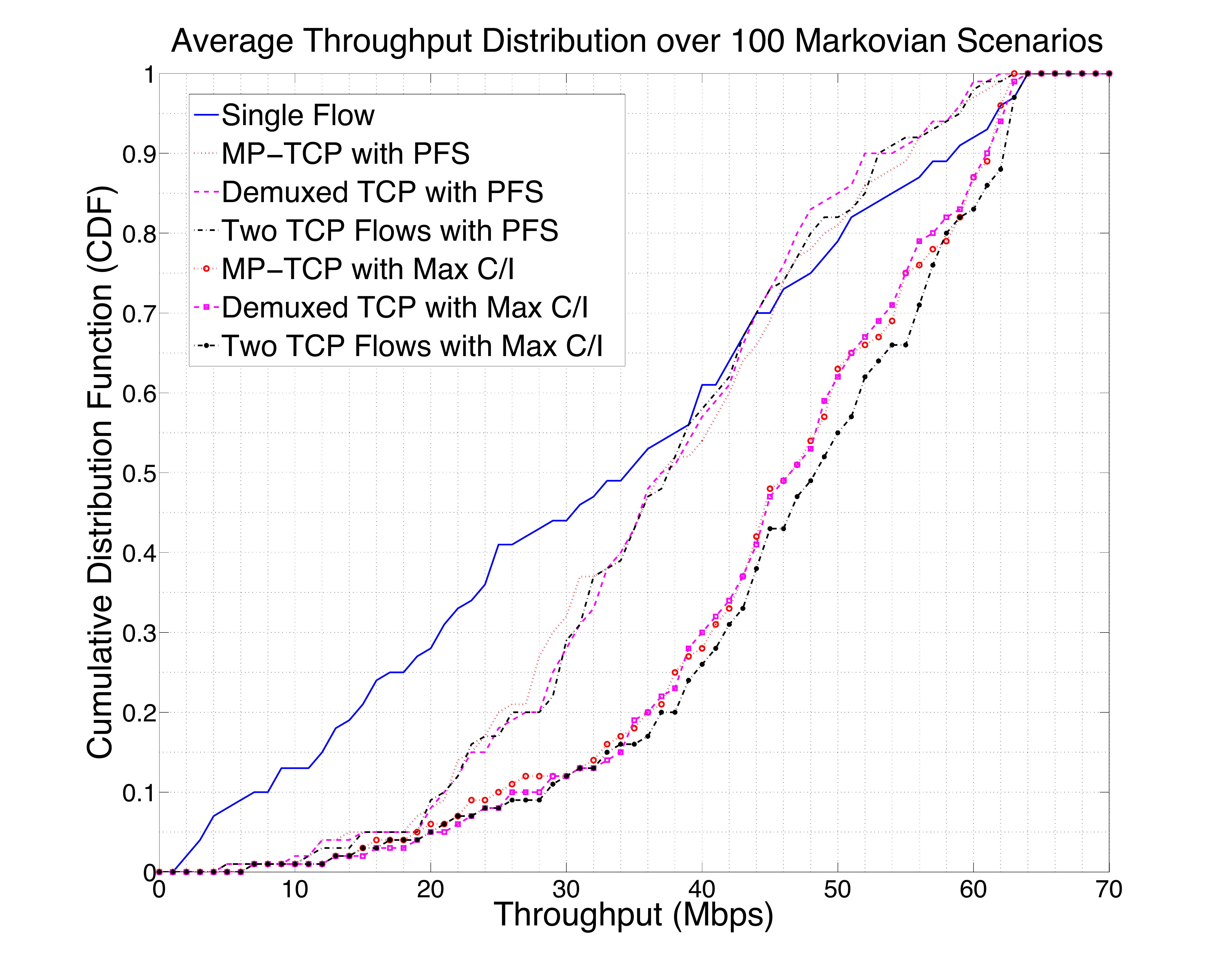}
	\caption{Throughput results for 100 randomly generated Markovian channel models for the scenario in Fig.~\ref{fig:mininet-topology}.}
	\label{fig:markov} 
	\vspace{-0.13in}
\end{figure}

For Markovian channel models, we generated 100 random scenarios (i.e., $p$ and $q$ values are set uniformly random in interval $(0,1)$ for each LTE device in each scenario). We run iperf3 to generate 5 minute long sessions using TCP-Cubic and MP-TCP distributions in Linux.  We report the cumulative distribution of average session throughput over 100 scenarios in  Fig.~\ref{fig:markov}.  In the figure, the curves that use PFS are representative of the cases that show what we can achieve today without any network programmability by completely relying on end to end solutions such as MP-TCP, multiple TCP connections and tethering. Against the baseline single flow case, they show significant improvement for half of the scenarios. But, for about 30\% of the scenarios where especially the relay node does not observe as good LTE channel conditions as the end host, these options without network intelligence delivers even worse throughput than the single flow case. On the other hand, when the network knows that both flows share the same eNB and alters the scheduling policy to Max C/I, we see that all multi-path transport options benefit from this (rightmost three curves). Naturally, the two TCP flow case with Max C/I performs the best as there is no overhead of flow splitting. Despite its overhead in flow splitting, Demuxed TCP actually performs a par with MP-TCP and quite comparable to the two TCP flow case. Thus, for legacy TCP applications, Demuxed TCP can deliver $3.5\times$ better throughput for the bottom 10\%, $2\times$ better performance for the bottom 30\%, and 34\% better performance for the median. Against the state of the art (MP-TCP with PFS), the respective gains are up 34\% for the bottom 10\% and  23\% for the median. 

\comment{Bottom 10\%  PFS 2.6x Max C/I   3.5x  Bottom 20\%   1.9x    2.4x Bottom 30\%    1.5x    2x  Bottom 40\%   1.4x    1.8x  Median   1.09x   1.34x}

\begin{figure}[!t]
	\centering
	\includegraphics[width=0.8\columnwidth]{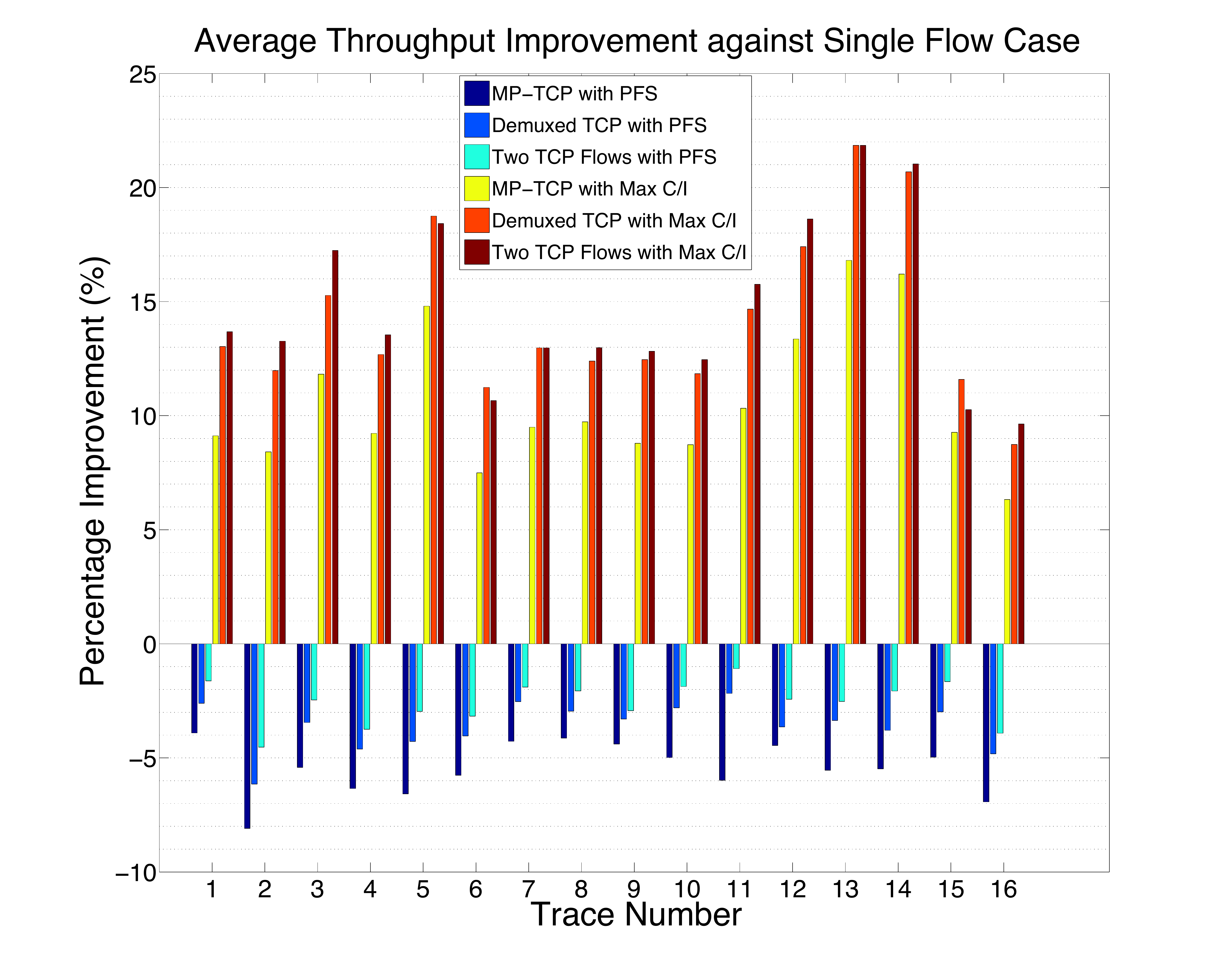}
	\caption{Throughput improvements over single flow case with real traces for the scenario in Fig.~\ref{fig:mininet-topology}.}
	\label{fig:realhigh} 
	\vspace{-0.13in}
\end{figure}

Since we have more limited number of real traces, instead of providing CDF curves, we summarize the results per scenario in Fig.~\ref{fig:realhigh} relative to the baseline single flow case. The single flow case achieves an average throughput in the range of $[18.4, 34.7]$Mbps for 16 scenarios with mean throughput of 29.1 Mbps across the scenarios. In all scenarios, without the network programmability and intelligence in the scheduling layer, all multi-path options fail to achieve better performance than the baseline observing up to 8\% performance loss! MP-TCP performs the worst in general. With programmability in scheduling, the picture is rectified and indeed significant improvements are achieved against the base line (up to 22\%). Given the limited number of traces we have, it is premature to conclude that multi-path diversity cannot be attained without programmability in the scheduling layer under shared wireless bottleneck. Nevertheless, this result is consistent with the Markovian channel models, where using multi-path routing without programmable scheduler performed worse than the baseline in high quality channel environments.

\begin{figure}[!t]
	\centering
	\includegraphics[width=0.8\columnwidth]{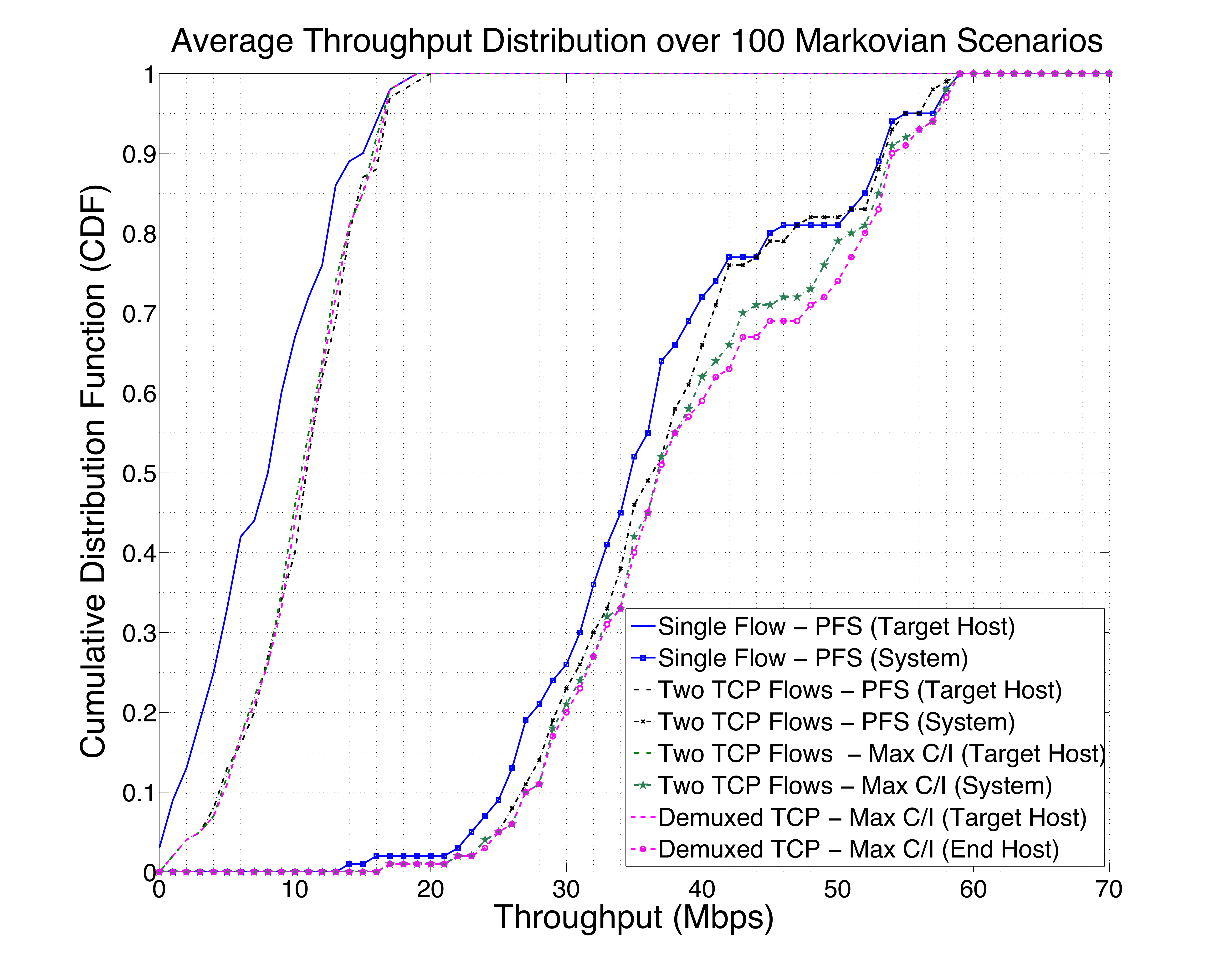}
	\caption{Targeted user and system throughput over Markovian channel models with additional destination UEs.}
	\label{fig:markovmu} 
	\vspace{-0.13in}
\end{figure}

\begin{figure}[!t]
	\centering
	\includegraphics[width=0.8\columnwidth]{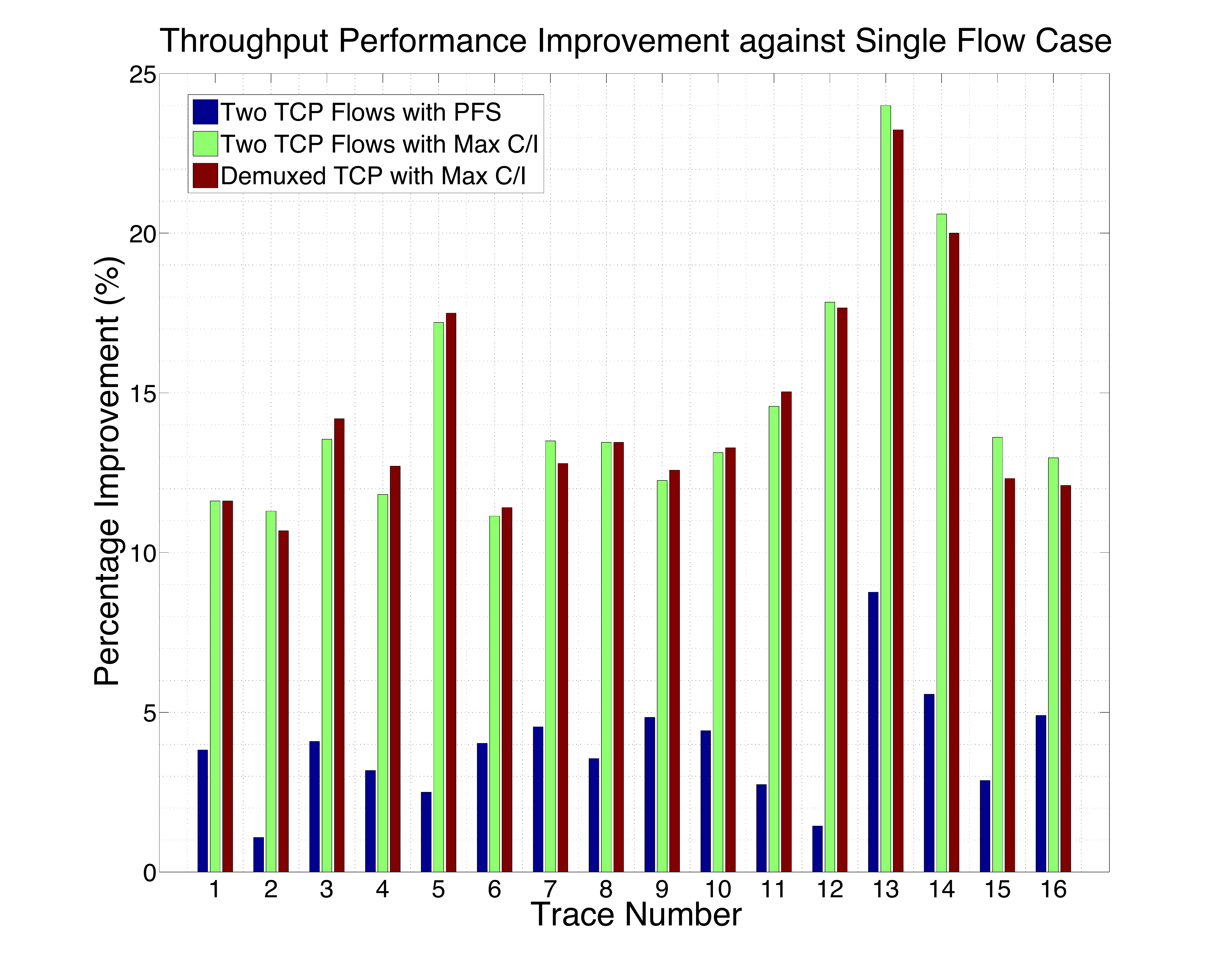}
	\caption{Targeted user throughput improvements over single flow case with real traces \& additional destination UEs.}
	\label{fig:realmu} 
	\vspace{-0.13in}
\end{figure}

We also added more users into the system to see how the multi-path diversity gains of a target host and the system throughput vary as we add more end hosts under the same LTE base station. Fig.~\ref{fig:markovmu} and Fig.~\ref{fig:realmu} present the performance results when three more destination nodes (with no D2D link) are added in  the topology in Fig.~\ref{fig:mininet-topology}. We omit the case for MP-TCP as it performs worse than the other multi-path options. For Markovian channel models, programmability in scheduling not only improves the throughput of user with D2D link (\emph{target host}) substantially,  but also improves the overall system throughput (i.e., sum rate across destination hosts). Demuxed TCP has a slight edge over the two TCP flow case with Max C/I. Without network programmability, the two TCP flow case pays a significant penalty in the system rate as it penalizes other end hosts that do not have the multi-path advantage. The evaluations over the real traces indicate similar findings: With programmability in the scheduling layer, the throughput of target host increases significantly (as much as 24\%). Although not shown in the figure, we also see that there is no rate reduction in other hosts (in fact they observe slight improvements). Without such programmability, the improvements for targeted host is more limited (less than 9\%) while the system throughput is reduced in some cases.   

\section{Conclusion}\label{sec:conclusion}
\comment{Due to the plethora of digital mobile devices,
the great coverage of LTE and the availability of high data rates,
more and more users are using the LTE in order to access the Internet.}
We argue that network programmability will be very critical in dense wireless environments
that include mobile as well as stationary wireless relays. Our prototyping over real transport protocols and 
applications show that the state of the art end to end solutions fail to take advantage
of the multi-user diversity over shared wireless resources. They even lead to performance losses in certain cases. Our evaluations utilize both synthetic channel models as well as real-time channel traces collected in a broadband LTE coverage area.  In contrast,  a simple programmability in the scheduling and routing layers
with the knowledge about which application flows share the same bottlenecks as well as the network topology including the D2D links
improves the user and system throughput substantially.


\end{document}